\documentclass[floatfix,12pt,prd,onecolumn,
superscriptaddress,nofootinbib]{revtex4}
\usepackage{latexsym}
\usepackage{epsfig}
\usepackage{amsmath}
\usepackage{epstopdf}
\usepackage{amssymb}
\usepackage{graphicx}
\usepackage{hyperref}
\usepackage{tabularx}
\usepackage{varioref}

\newcommand{\be}{\begin{equation}}
\newcommand{\ee}{\end{equation}}
\newcommand{\bea}{\begin{eqnarray}}
\newcommand{\eea}{\end{eqnarray}}

\usepackage[utf8]{inputenc}
\begin{document}

\title{PV criticality of Achucarro-Ortiz black hole in the presence of
higher order quantum and GUP corrections}
\author{Behnam Pourhassan}
\email{b.pourhassan@candqrc.ca}
\affiliation{Iran Science Elites Federation, Tehran, Iran}
\affiliation{Canadian Quantum Research Center 204-3002 32 Ave Vernon, BC V1T 2L7 Canada}
\author{Ali \"{O}vg\"{u}n}
\email{ali.ovgun@pucv.cl}
\affiliation{Instituto de F\'{\i}sica, Pontificia Universidad Cat\'olica de Valpara\'{\i}%
so, Casilla 4950, Valpara\'{\i}so, Chile.}
\affiliation{Physics Department, Arts and Sciences Faculty, Eastern Mediterranean
University, Famagusta, North Cyprus via Mersin 10, Turkey.}
\author{\.{I}zzet Sakall{\i}}
\email{izzet.sakalli@emu.edu.tr}
\affiliation{Physics Department, Arts and Sciences Faculty, Eastern Mediterranean
University, Famagusta, North Cyprus via Mersin 10, Turkey.}
\date{\today}

\begin{abstract}
In this paper, we study quantum-corrected and GUP corrected thermodynamics of the (2+1)-dimensional
charged-rotating Achucarro-Ortiz black hole. The corrected parameters
include temperature, entropy, and heat capacity which help to investigate
the instability phases of the Achucarro-Ortiz black hole. We show that this black hole with
small mass possesses unstable regions. However, we reveal that those
instabilities can be removed by the GUP corrections. Finally, we also
compute the maximum temperature that can be reached by the Achucarro-Ortiz black hole. We show that corrected temperatures by different methods we used are identical at small mass limit, hence GUP correction at small mass limit is a quantum gravity correction. Interestingly, we show by graphical analysis that leading order quantum corrected temperature of the Achucarro-Ortiz black hole behaves similar to the GUP corrected temperature of uncharged Achucarro-Ortiz black hole.
\end{abstract}

\maketitle

\newpage

\section{Introduction}
Banados-Teitelboim- Zanelli (BTZ) black hole \cite{BTZ} is one of the
important kinds of black holes in three-dimensional space-time. The rotating
charged BTZ black hole solution was studied by Achucarro and Ortiz \cite{AO}, which is called the AO (Achucarro-Ortiz) black holes. It has been found that entropy of Achucarro-Ortiz black hole can be described by
the Cardy-Verlinde formula \cite{setare}. Also, Hawking radiation of the AO black hole discussed by the Ref. \cite{X}. Indeed, Kaluza-Klein reductional dimension help to obtain lower dimensional black holes \cite{kk1,kk2}. Since then, the AO black holes have been taking attention both in
theoretical physics and cosmology \cite{A1,A2,A3,A4}. For example, the
effects of quantum fluctuations (first order) on the properties of a charged
BTZ black hole in massive gravity were studied in the Ref. \cite{A3}.
Recently, rotating BTZ black hole with higher order corrections of the
entropy has been investigated and shown that it elicits some instabilities
\cite{1}. Besides, it was seen that when a logarithmic correction is
considered for the uncharged BTZ black hole, the leading-order corrections
yield an instability while the higher order corrections remove them \cite{1-2}. It is worth noting that thermodynamics of higher dimensional black
holes with higher order thermal fluctuations have been studied in \cite{3-0}. These kinds of black holes are important to study AdS/CFT correspondence in lower dimensions.\\
In the present study, our main goal is to study the thermodynamics of the AO
black hole with higher order quantum corrections. In particular, we study how
the correction terms affect the stability of the AO black hole. Correction
terms include logarithmic one \cite{das, SPR} together with the higher order
term, which is proportional to the inverse of the entropy \cite{more}. These
correction terms indeed arise from the thermal fluctuations of the
statistical physics, which may be interpreted as quantum corrections when
the size of the black hole becomes infinitesimal \cite{NPB}. Thermal
fluctuations are important in several backgrounds like a hyperscaling
violation background \cite{EPJC33}. By using the small black holes
including the quantum effects, one can consider the quantum gravity effects \cite{Annals, NPB2} (see also \cite{ex1,ex2,2,3,Feng1,Feng2,Feng3}). Moreover, the logarithmic and higher order corrections are considered to study the critical
thermodynamics behaviors of some black objects like a dyonic charged AdS
black hole \cite{PRD}, charged dilatonic black Saturn \cite{sat1,sat2}, and
AdS black holes in massive gravity \cite{sudb}. In that case, it is possible
to have the holographic dual of a Van der Waals fluid \cite{Kubiznak:2012wp,Gunasekaran:2012dq,Kubiznak:2016qmn,Kubiznak:2014zwa}. Hence, we shall
investigate the $P$-$V$ diagram of the AO black hole to find such a Van der Waals behavior in the presence of higher order corrections \cite{Anabalon:2018ydc,Zou:2017juz,Zou:2014gla,Zhang:2014eap,Ma:2017pap,Ma:2016lwr,Zhao:2014raa,Okcu:2016tgt,Hendi:2018sbe,Momennia:2017hsc,Hendi:2015eca}.
On the other hand, the emergence of a minimal observable distance yields to the generalized uncertainty principle (GUP) \cite{GUP1, GUP22, GUP3}, which may be affect the black hole temperature and entropy. One of the key frontiers in modern theoretical physics is to construct a renormalizable, UV complete and non-perturbative theory of quantum gravity which could explain the features near the singularity of black hole and the Big Bang. Although numerous candidates of such theories are proposed, however, most of them offer no testable predictions or even untestable experimentally.
In Ref. \cite{Giovanazzi}, (1 + 1)-D black hole entropy investigated by the brick-wall method. The other idea to solve the divergences is to consider the modified Heisenberg uncertainty relation which shows that there exists a minimal length. Thus, using the modified Heisenberg uncertainty relation the divergence in the brick-wall model are eliminated. In the absence of experiments, thermodynamics offer us a physically acceptable route to understand strong gravity regimes. The semi-classical physics help to identifying the properties of black hole like area and surface gravity with thermodynamical features such as entropy and temperature. Any quantum gravity (QG) theory should offer additional terms or correction terms to the results of semi-classical physics. In this connection, one of the aims of 2+1 dimensional QG is construct toy models of black holes and analyze their thermal properties. Since GUP and log corrections are motivated by numerous QG theories, it is imperative to apply these corrections to lower space dimensional black holes. These corrections play prominent role at the smaller scales closer to the Planck scale. Hence, we would like to consider such quantum effect and study modified thermodynamics in AO black hole.\\
The plan of the paper is organized as follows: In section 2, we briefly review
the AO black hole space-time and its corrected thermodynamics due to the thermal fluctuations. We use the first and second order corrected black hole entropy to show the effects of thermal fluctuations (which interpreted as quantum effects) on the AO black hole thermodynamics quantities and its stability. In this section we assume that the black hole temperature not affected by quantum corrections. In section 3, we compute
the higher order quantum corrected temperature of the AO black hole. Section
4 is devoted to the GUP corrected entropy and temperature of AO black hole.
In section 5, we summarize our results and discuss about relations between different corrections.

\section{Achucarro-Ortiz black hole}
\begin{figure}[h!]
\begin{center}
$%
\begin{array}{cccc}
\includegraphics[width=70 mm]{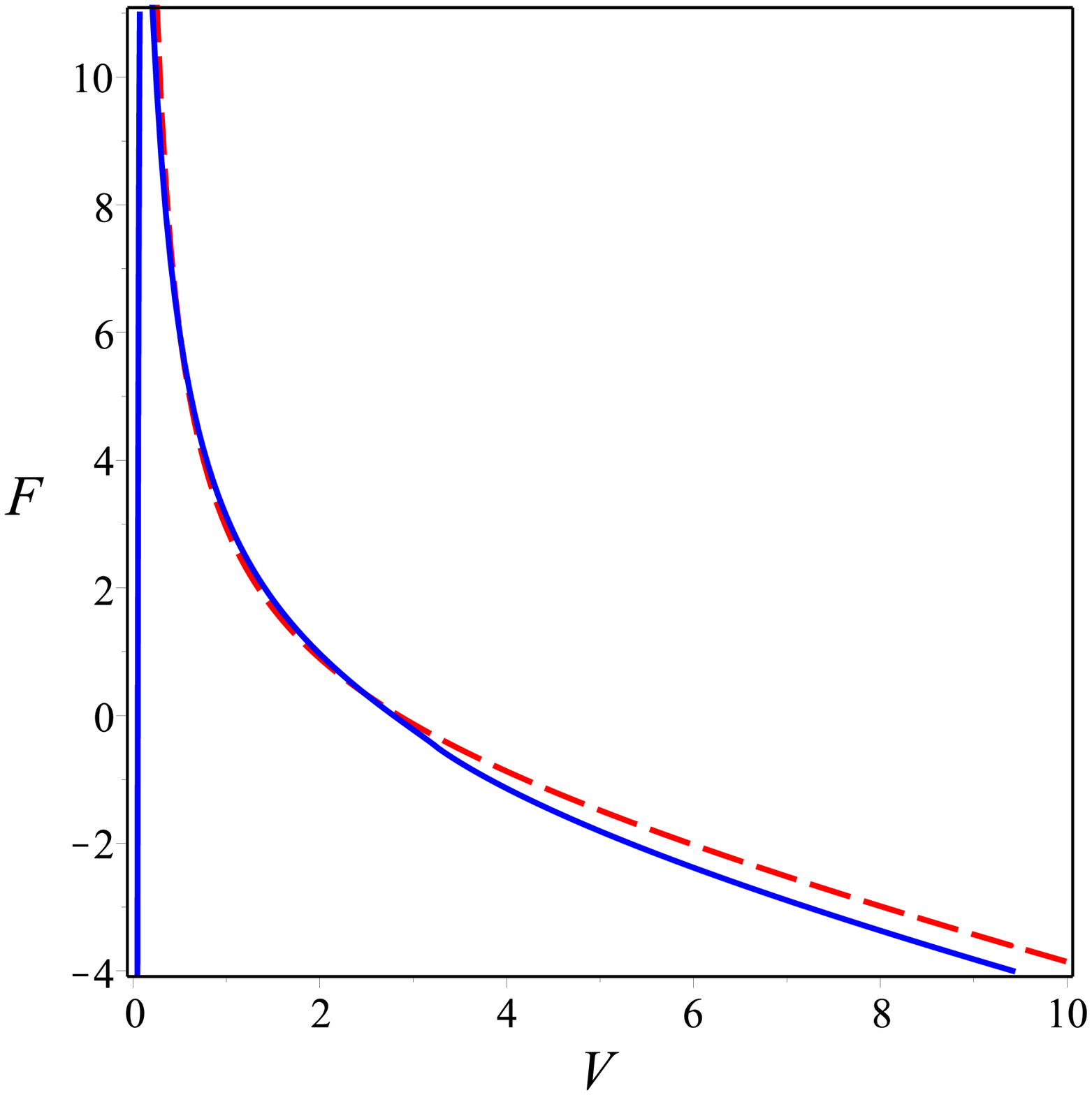}&\includegraphics[width=70 mm]{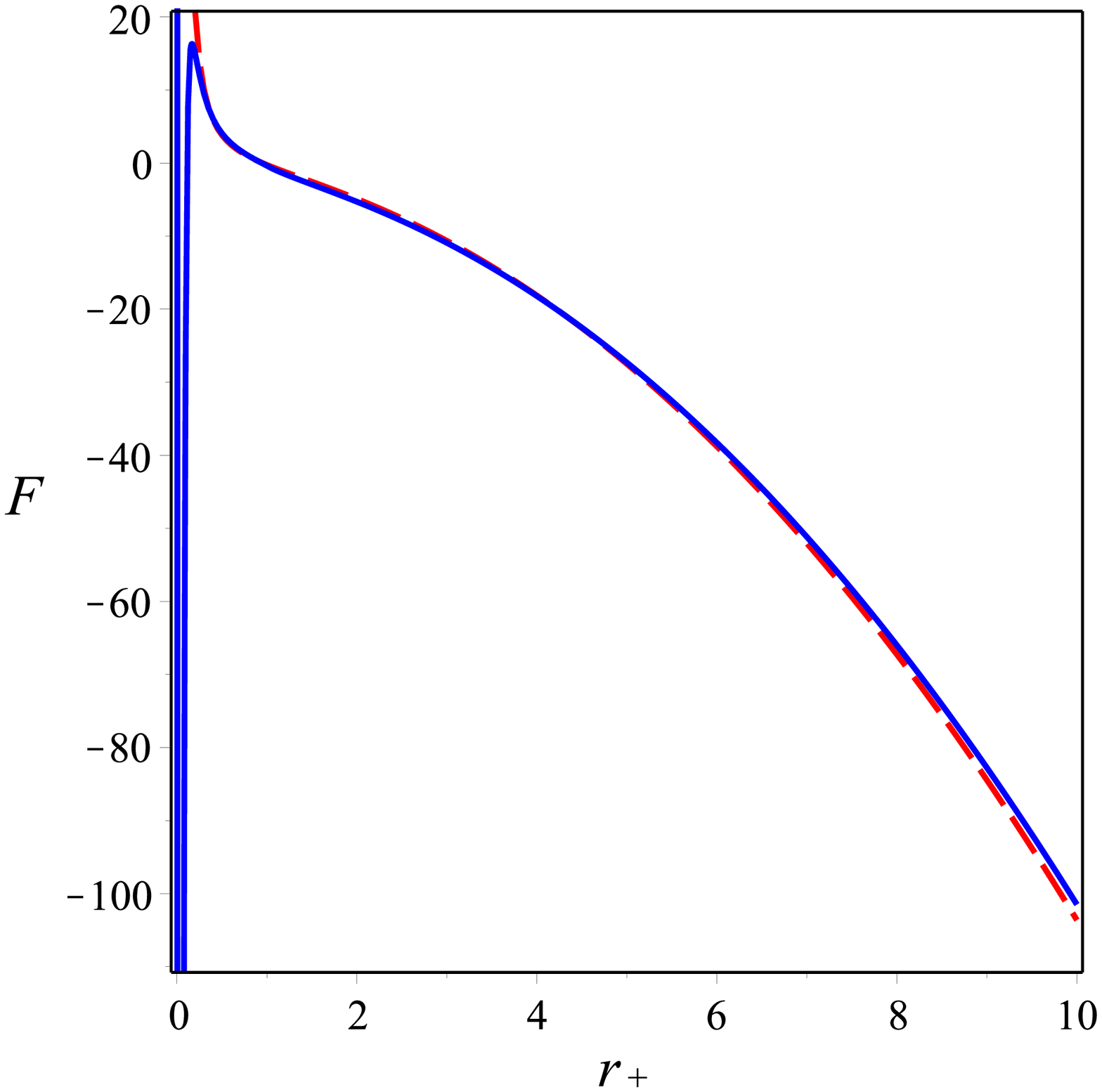}
\end{array}%
$%
\end{center}
\caption{Helmholtz free energy in terms of $V$ (left plot) and in terms of $%
r_{+}$ (right plot) with $Q=J=l=1$. Dashed red lines represent the ordinary
case with $\protect\alpha=\protect\beta=0$. Solid blue lines represent
corrected case with $\protect\alpha=\protect\beta=1$.}
\label{fig:1}
\end{figure}

The Einstein-Maxwell action in $2+1$ dimensions is given by,
\begin{equation}
I=\int d^{3}x\sqrt{-g}\Big(R-2\Lambda-\frac{1}{4}F_{ab}F^{ab}\Big).
\end{equation}
In the case of $\Lambda=0$ we recover the action given by the Ref. \cite{4002}.
The Einstein field equations for ($2+1$)-dimensional space-time with
negative cosmological constant take the following form
\begin{equation}
G_{ab}+\Lambda g_{ab}=\pi T_{ab}~~~~~~~(a,b=0,1,2),
\end{equation}%
which results in the BTZ black hole solution with electric charge and spin:
\begin{equation}
ds^{2}=-f(r)dt^{2}+\frac{dr^{2}}{f(r)}+r^{2}\Big(d\phi-\frac{J}{2r^{2}}%
dt\Big)^{2}.  \label{metric}
\end{equation}%
The above line-element is also called AO black hole \cite{AO}. In Eq. (\ref%
{metric}), the metric function $f(r)$ reads
\begin{equation}
f(r)=-M+\frac{r^{2}}{l^{2}}+\frac{J^{2}}{4r^{2}}-\frac{\pi }{2}Q^{2}\ln {\frac{r}{l}},
\end{equation}
where $M$, $Q$, $J$ denote the mass, electric charge, and angular momentum
of the black hole, respectively. Also, $\Lambda =-1/l^{2}$ is the negative
cosmological constant and $l$ is the AdS length. As we know the holographic quantity of pressure in the AdS black holes is the cosmological constant. Therefore, we can write,
\begin{equation}\label{pressure}
P=\frac{1}{16\pi l^{2}}.
\end{equation}
The event horizon (or the stationary limit surface)
which is a null hypersurface of this black hole occurs when $g^{rr}=0$:
\begin{equation} \label{mass}
-M+\frac{r_{+}^{2}}{l^{2}}+\frac{J^{2}}{4r_{+}^{2}}-\frac{\pi }{2}Q^{2}\ln{\frac{r_{+}}{l}}=0,
\end{equation}%
which yields the mass of the black hole as
\begin{equation}\label{mass2}
M=\frac{4r_{+}^{4}-2\pi Q^{2}l^{2}r_{+}^{2}\ln{\frac{r_{+}}{l}}+J^{2}l^{2}}{4l^{2}r_{+}^{2}}.
\end{equation}
The Hawking temperature of the black hole is given by
\begin{equation}\label{temp}
T_{H}=\frac{1}{4\pi }\left. \frac{\partial f(r)}{\partial r}\right\vert
_{r=r_{+}}=\frac{1}{4\pi }\Big(\frac{2r_{+}}{l^{2}}-\frac{J^{2}}{2r_{+}^{3}}-%
\frac{\pi Q^{2}}{2r_{+}}\Big).
\end{equation}%
The entropy is associated with the event horizon as $S_{0}=4\pi r_{+}$. The
thermodynamic volume is $V=\pi r_{+}^{2}$, which means that event horizon
can be expressed as $r_{+}=\sqrt{\frac{V}{\pi }}$.\\
In that case the first law of thermodynamics reads as,
\begin{equation}\label{first}
dM=TdS_{0}+\Omega dJ+\Phi dQ+V dP,
\end{equation}
where
\begin{equation}\label{Omega}
\Omega=\left(\frac{dM}{dJ}\right)_{S_{0}, Q, P}=\frac{J}{2r_{+}^{2}},
\end{equation}
is the horizon angular velocities while
\begin{equation}\label{Pot}
\Phi=\left(\frac{dM}{dQ}\right)_{S_{0}, J, P}=-\pi Q\ln{\frac{r_{+}}{l}},
\end{equation}
is the horizon electrostatic potential.\\
The logarithmic corrected entropy expression (or the first order correction)
is given by \cite{SPR},
\begin{equation}\label{myent}
S=S_{0}-\frac{\alpha }{2}\ln (S_{0}T_{H}^{2}),
\end{equation}
where the constant $\alpha$ is added to track the correction term \cite{Pd,EPL}. In that case if we choose $\alpha=0$, the expression for the
entropy without any corrections recovered. Moreover, in the case of $\alpha=1$, we obtain the corrections due to thermal fluctuations. Hence, for the large black holes in low temperature, we can take the limit $\alpha\rightarrow0$, and for the small black holes in high temperature, we can
take the limit $\alpha\rightarrow1$.
In general, one can consider arbitrary $\alpha$ and fix it by
thermodynamics requirements or observational data in higher dimensions. One can find out the explicit form of Eq. (\ref{myent})
as follows
\begin{equation}
S=4\pi r_{+}-\frac{1}{2}\alpha \Big[-4\ln 2-\ln \pi +\ln \Big(\frac{(\pi
l^{2}Q^{2}r_{+}^{2}+J^{2}l^{2}-4r_{+}^{4})^{2}}{l^{4}r_{+}^{5}}\Big)\Big].
\end{equation}%
The Helmholtz function is given by
\begin{equation}
F=-\int SdT_{H}=-\int S(r_{+})\frac{dT_{H}}{dr_{+}}dr_{+},
\end{equation}%
which yields
\begin{equation*}
F=\frac{1}{\pi r_{+}^{3}l^{2}}\Big(\frac{-1}{2}\pi
^{2}Q^{2}r_{+}^{3}l^{2}\ln{\frac{r_{+}}{l}}+\frac{3}{4}\pi r_{+}l^{2}J^{2}-\pi r_{+}^{5}%
\Big)+F_{1}(\alpha ),
\end{equation*}%
where $F_{1}(\alpha )$ is a long expression of the first correction term. We
now proceed to the second order correction to the entropy as follows:
\begin{equation}\label{Gentropy}
S_{c}=S_{0}-\frac{\alpha }{2}\ln (S_{0}T_{H}^{2})+\frac{\beta }{S_{0}},
\end{equation}%
where the constant $\beta$ shows the higher order correction.
In general, we know that all different approaches to quantum gravity in leading order generate the
logarithmic corrections to the black hole entropy while inverse of entropy in higher order. We should note that although
the leading order correction is logarithmic, but the
coefficient is depends on the quantum gravity theory, hence we one can consider this coefficient as a free parameter of the model. Since
the values of the coefficients depend on the quantum
gravity approach, it can be argued that they are generated from quantum fluctuations of the space-time
geometry rather than matter fields on space-time. Therefore, we consider general $\alpha$ and $\beta$ to obtain modified thermodynamics. In that case the corrected first law of thermodynamics,
\begin{equation}\label{cfirst}
dM=TdS_{c}+\Omega dJ+\Phi dQ+V dP,
\end{equation}
may be violated which show some instabilities in presence of $\alpha$ and $\beta$ which will be discussed later. However, we can consider these coefficients like Lagrange multiplier and fix them simultaneously to satisfy the corrected first law of thermodynamics. It means that the equation (\ref{cfirst}) yields to the following equation,
\begin{equation}\label{multiplier}
\left(\frac{\alpha}{2}+\frac{\beta}{S_{0}}\right)dS_{0}+\alpha\frac{S_{0}}{T}dT=0.
\end{equation}
The corrected entropy (\ref{Gentropy}) yields the following relevant correction term to the
Helmholtz function:
\begin{equation}
F_{c}=F+\beta \Big(\frac{Q^{2}}{64\pi r_{+}^{2}}+\frac{3J^{2}}{128\pi
^{2}r_{+}^{4}}-\frac{\ln{\frac{r_{+}}{l}}}{8\pi ^{2}l^{2}}\Big).
\end{equation}%
In Fig. \ref{fig:1}, one can see the typical behavior of the Helmholtz free
energy for the corrected and uncorrected cases. We see some infinitesimal
variation in the Helmholtz function as being observed in \cite%
{Kuang:2018goo,Ovgun:2017bgx}. It can be also seen that there is a maximum
value for the Helmholtz free energy. The correction terms reduce the maximum
value.\\

\begin{figure}[h!]
\begin{center}
$%
\begin{array}{cccc}
\includegraphics[width=90 mm]{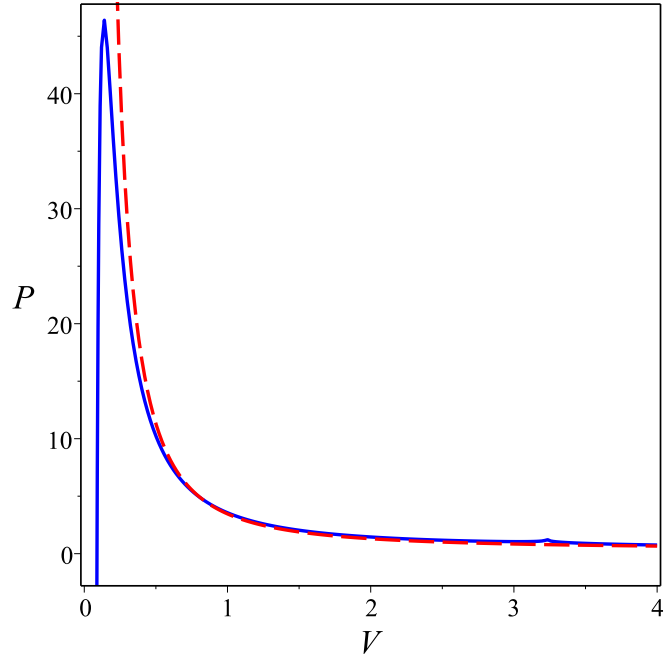}
\end{array}%
$%
\end{center}
\caption{Pressure in terms of $V$. Dashed red lines represent the
ordinary case with $\protect\alpha=\protect\beta=0$, while solid blue lines
represent corrected case with $\protect\alpha=\protect\beta=1$.}
\label{fig:2}
\end{figure}

To find the equation of state, we employ the first derivative of $F_{c}$ with respect
to volume, more specifically:
\begin{equation}
P=-\frac{\partial F_{c}}{\partial V},
\end{equation}%
which gives
\begin{equation*}
P=\frac{\pi^{2}Q^{2}Vl^{2}+3J^{2}\pi^{2}l^{2}+4V^{2}}{4\pi l^{2}V^{2}}+{\mathcal{O}}(\alpha )+{\mathcal{O}}(\beta ).
\end{equation*}%
To avoid cumbersome mathematical expressions like above, from now on we
shall write only the leading order term, however all terms will be taken
into account for the numerical purposes.\\
In Fig. \ref{fig:2}, one can deduce from the $P$-$V$ diagram that there is
no Van der Waals behavior as well as no critical points. When the size of
the black hole is at microscopic scale, the pressure becomes negative and
black hole exhibits unstable phases. On the other hand, logarithmic
and higher order corrections yield instability for the small black hole. In
other words, when the black hole size is reduced by the Hawking radiation,
thermal fluctuations of the quantum corrections become important and their
effects cause to the instability.\\

To obtain internal energy, we use the following thermodynamic relation:
\begin{equation}
E=F_{c}+S_{c}T_{H},
\end{equation}%
which gives us the following expression
\begin{equation}
E=F+\beta \Big(\frac{Q^{2}}{64\pi r_{+}^{2}}+\frac{3J^{2}}{128\pi
^{2}r_{+}^{4}}-\frac{\ln{\frac{r_{+}}{l}}}{8\pi ^{2}l^{2}}\Big)+S_{0}T_{H}-\frac{\alpha
}{2}T\ln (S_{0}T_{H}^{2})+\frac{\beta }{S_{0}}T_{H}.
\end{equation}%
Further, we can calculate enthalpy
\begin{equation}
h=E+PV,
\end{equation}%
and the Gibbs free energy:
\begin{equation}
G=H-T_{H}S_{c}.
\end{equation}%
Our numerical analysis show that the behavior of enthalpy and Gibbs energy
are similar to the internal energy.\\
\begin{figure}[h!]
\begin{center}
$%
\begin{array}{cccc}
\includegraphics[width=80 mm]{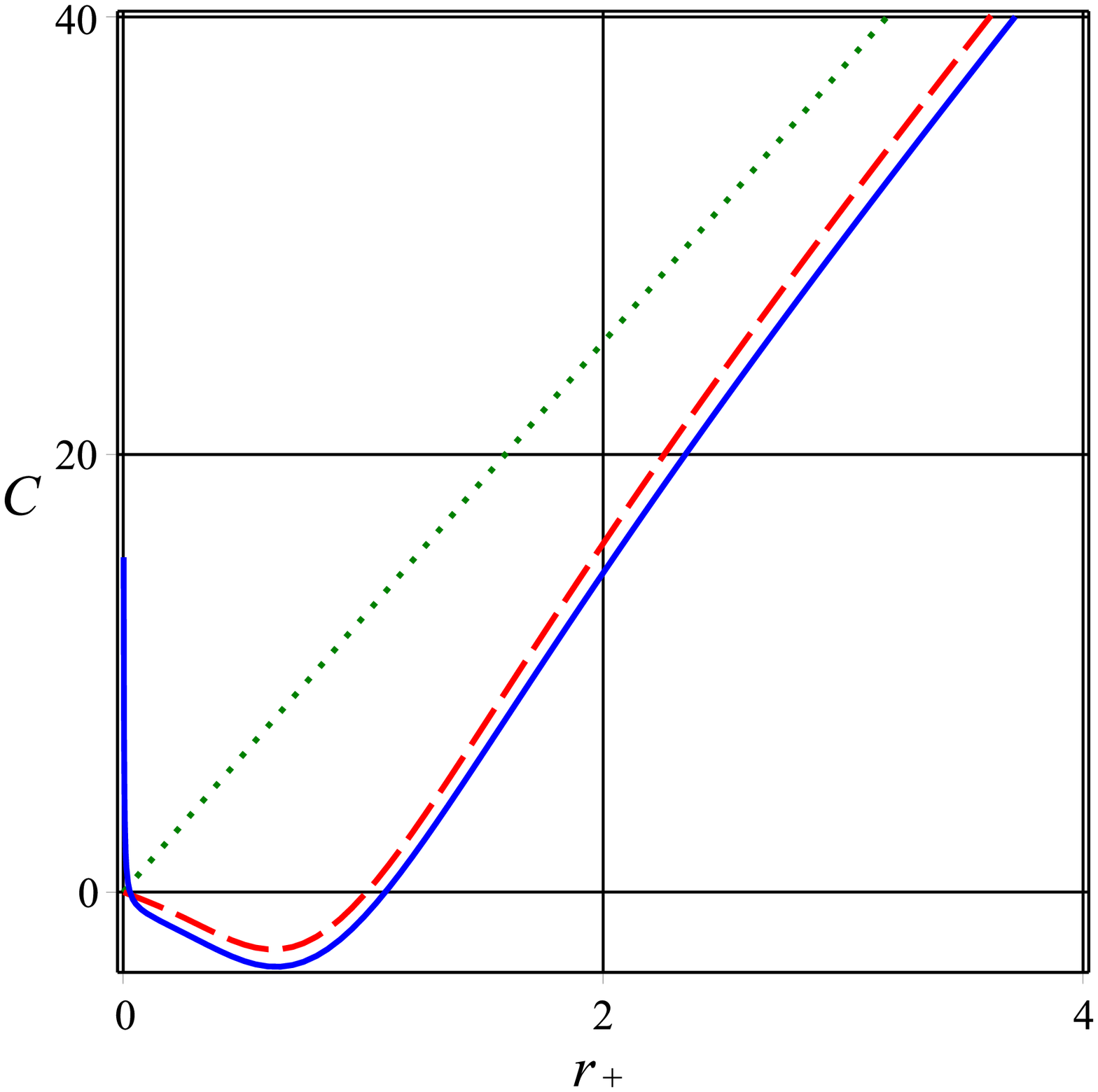}
\end{array}%
$%
\end{center}
\caption{Specific heat in terms of $r_{+}$. Dotted green line
represent the uncharged static case ($Q=J=0$), dashed red lines represent
the ordinary case with $\protect\alpha=\protect\beta=0$ and $Q=J=1$, while
solid blue lines represent corrected case with $\protect\alpha=\protect\beta%
=1$ and $Q=J=1$.}
\label{fig:3}
\end{figure}
To investigate the stability of the physical system, we consider the
specific heat definition:
\begin{equation}
C=T_{H}\frac{\partial S_{c}}{\partial T_{H}}
\end{equation}%
which gives
\begin{equation}
C=4\pi \frac{\frac{2r_{+}}{l^{2}}-\frac{J^{2}}{2r_{+}^{3}}-\frac{\pi Q^{2}}{%
2r_{+}}}{\frac{2}{l^{2}}+\frac{3J^{2}}{2r_{+}^{4}}+\frac{\pi Q^{2}}{%
2r_{+}^{2}}}+C_{1}(\alpha )+C_{2}(\beta ),
\end{equation}%
where $C_{1}(\alpha )$ and $C_{2}(\beta )$ are complicated $\alpha $ and $%
\beta $ dependent terms. We shall make their physical interpretation,
graphically. Sign of the specific heat and its asymptotic behavior can give
us information about the stability and phase transition \cite{EPJC22}.\\
In Fig. \ref{fig:3}, we draw the specific heat versus event horizon graph.
In the case of uncharged static AO black hole (dotted green line), we see
completely stable black hole. However, with the inclusion of rotation, some
instabilities appear for small radii. In the presence of correction terms,
unstable regions increased. Although there are some unstable phases, however
there is no critical point and phase transition corresponding to asymptotic
behavior.

\section{Higher Order Quantum Corrected Temperature}
In this section, we shall attempt to derive higher order quantum corrected
temperature of BTZ black hole, which is dual of one dimensional holographic
superconductors \cite{Refiz1}. To this end, we shall study the
Parikh-Wilczek's quantum tunneling method \cite{Refiz2} together with the
entropy (\ref{Gentropy}) to add higher order quantum corrections to the
tunneling probability by considering the back reaction effects. However, the effects of Heisenberg uncertainly principle i.e., GUP corrections will be discussed in the next section. Finally, the
modified $T_{H}$ due to the back reaction effect will be computed. However,
as can be seen from the operations that are detailed below, it is necessary
to express the entropy in terms of mass to be able to do all of these. For
this, it is necessary to write the horizon in terms of mass. But, the
transcendental structure of Eq. (\ref{mass2}) does not allow this. To
overcome this difficulty, we simply consider the chargeless case: $Q=0$.
Thus, from Eq. (\ref{mass2}), one can get event and inner horizons as follows,
\begin{equation}
r_{+}=\frac{1\,}{\sqrt{2}}\sqrt{M{l}^{2}+l\sqrt{{M}^{2}{l}^{2}-{J}^{2}}},
\label{iz1}
\end{equation}
and
\begin{equation}
r_{-}=\frac{1\,}{\sqrt{2}}\sqrt{M{l}^{2}-l\sqrt{{M}^{2}{l}^{2}-{J}^{2}}},
\label{iz1n}
\end{equation}
such that we have
\begin{equation}
M=\frac{r_{+}^{2}-r_{-}^{2}}{l},  \label{iz1n2}
\end{equation}
and
\begin{equation}
J=\frac{2r_{+}r_{-}}{l},  \label{iz1n3}
\end{equation}
and also the Hawking temperature (\ref{temp}) becomes
\begin{equation}
T_{H}=\frac{1}{4\pi }\left( \frac{2r_{+}}{l^{2}}-\frac{J^{2}}{2r_{+}^{3}}%
\right) .  \label{iz0m}
\end{equation}
The coordinate system in Eq. (\ref{metric}) is described for the observer
located at spatial infinity. After transforming the metric (\ref{metric}) to
the dragging coordinate system \cite{Refiz3}:
\begin{equation}
\psi =\phi -\frac{J}{2r^{2}}t,\text{ \ \ } \mathcal{T}=t\text{,}
\label{iz1n4}
\end{equation}%
we see that the physics near the horizon can be effectively ($\psi =const.$)
described by the following two-dimensional metric
\begin{equation}
ds^{2}=-f(r)d\mathcal{T}^{2}+\frac{dr^{2}}{f(r)}.  \label{iz1n5}
\end{equation}
On the other hand, the near horizon metric (\ref{iz1n5}) can be expressed in
the regular Painleve-Gulstrand (PG) coordinates \cite{Refiz4,Refiz5} by
applying the following transformation
\begin{equation}
d\mathcal{T}_{PG}=d\mathcal{T}+\frac{\sqrt{1-f(r)}}{f(r)}dr,  \label{iz1n6}
\end{equation}
where $\mathcal{T}_{PG}$ is called the PG time, which is nothing but the
proper time. Thus, the metric (\ref{iz1n5}) recasts in
\begin{equation}
ds^{2}=-f(r)d\mathcal{T}_{PG}^{2}+2\sqrt{1-f(r)}d\mathcal{T}_{PG}dr+dr^{2},
\label{iz1n7}
\end{equation}
and it admits the following radial null geodesics of a test particle:
\begin{equation}
\dot{r}=\frac{dr}{d\mathcal{T}_{PG}}=-\sqrt{1-f(r)}\pm 1,  \label{iz1n8}
\end{equation}
where plus (minus) sign corresponds to outgoing (ingoing) geodesics. After
expanding Eq. (\ref{iz1n8}) around the event horizon, one can find that the
radial outgoing null geodesics, $\dot{r}$,\ is approximated to the following
\begin{equation}
\dot{r}\cong \kappa (r-r_{+}).  \label{izn9}
\end{equation}
in which the surface gravity \cite{Refiz6} (in the PG coordinates) reads
\begin{equation}
\kappa =\frac{1}{2}\left. \frac{\partial f(r)}{\partial r}\right\vert
_{r=r_{+}}.  \label{izn10}
\end{equation}
On the other hand, the imaginary part of the action ($I$) for an outgoing
particle with positive energy that crosses the event horizon from inside ($%
r_{\otimes }$) to outside ($r_{\odot }$) is given by \cite{Refiz7}
\begin{equation}
Im{I}=Im\int_{r_{\otimes }}^{r_{\odot }}p_{r}dr=Im\int_{r_{\otimes
}}^{r_{\odot }}\int_{0}^{p_{r}}d\tilde{p}_{r}dr.  \label{iz2}
\end{equation}
Let us recall that the Hamilton's equation for the classical trajectory is
given by
\begin{equation}
dp_{r}=\frac{dH}{\dot{r}},  \label{iz3}
\end{equation}
where $p_{r}$ and $H$\ are the radial canonical momentum and Hamiltonian,
respectively. Thus, we have
\begin{equation}
Im{I}=Im\int_{r_{\otimes }}^{r_{\odot }}\int_{0}^{H}\frac{d\widetilde{H}}{%
\dot{r}}dr.  \label{iz4}
\end{equation}
Now, let us assume that we have a circularly symmetric space-time having
constant total mass $M$. By making one more assumption, we consider the
system as if containing a radiating BTZ black hole with varying mass $%
M-\omega $ that emits a circular shell of energy $\omega $: $\omega \ll M$.
This scenario describes the self-gravitational effect \cite{Refiz8,Refiz9}.
In this framework, Eq. (\ref{iz4}) becomes
\begin{align}
ImI& =Im\int_{r_{\otimes }}^{r_{\odot }}\int_{M}^{M-\omega }\frac{d%
\widetilde{H}}{\dot{r}}dr,  \notag \\
& =-Im\int_{r_{\otimes }}^{r_{\odot }}\int_{0}^{\omega }\frac{d\widetilde{%
\omega }}{\dot{r}}dr,  \label{iz5}
\end{align}
in which the Hamiltonian $H=M-\omega $\ and $dH=-d\omega $ are used.
Following Eq. (\ref{izn9}), the radial outgoing null geodesics, $\dot{r}$,\
of the radiating black hole is defined as follows \cite{Refiz7}
\begin{equation}
\dot{r}\cong \kappa _{QGC}(r-r_{+}),  \label{iz6}
\end{equation}
where $\kappa _{QGC}=\kappa (M-\omega )$\ is the quantum gravity corrected ($%
QGC$) surface gravity \cite{Refiz10,Refiz11}. Therefore, after $r$
integration (the integration over $r$ is done by deforming the contour), Eq.
(\ref{iz5}) becomes
\begin{equation}
ImI=-\pi \int_{0}^{\omega }\frac{d\tilde{\omega}}{\kappa _{QGC}}.
\label{iz7}
\end{equation}
After defining the $QGC$ Hawking temperature as $T_{QGC}=\frac{\kappa _{QGC}%
}{2\pi }$, we get
\begin{equation}
ImI=-\frac{1}{2}\int_{0}^{\omega }\frac{d\tilde{\omega}}{T_{QGC}}=-\frac{1}{2%
}\int_{S_{QGC}(M)}^{S_{QGC}(M-\omega )}dS=-\frac{1}{2}\Delta S_{QGC.}
\label{iz8}
\end{equation}
The above expression yields the modified tunneling rate:
\begin{equation}
\Gamma _{QGC}\sim e^{-2ImI}=e^{\Delta S_{QGC}}.  \label{iz9}
\end{equation}
Taking cognizance of Eq. (\ref{Gentropy}), one can compute $\Delta S_{QGC}$
as follows
\begin{eqnarray}  \label{iz10}
\Delta S_{QG} &=&S_{QGC}(M-\omega )-S_{QGC}(M) \nonumber\\
&=&4\pi \left( r_{+}(\omega)-r_{+}\right)+\frac{\alpha }{2}\left[ \ln \left( \frac{r_{+}}{4\pi}\left( 2\,{l}^{-2}+{%
\frac{6{J}^{2}}{{r_{+}}^{2}}}\right) ^{2}\right) \right] \nonumber\\
&-&\left[ \ln \left( \frac{r_{+}(\omega)}{4\pi}\left( 2\,{l}^{-2}+{\frac{6{J}%
^{2}}{{r_{+}(\omega)}^{2}}}\right) ^{2}\right) \right]-\frac{\beta }{2\pi }\left( {\frac{1}{2r_{+}}}-{\frac{1}{2r_{+}(\omega)}}%
\right),
\end{eqnarray}
where $r_{+}$ is given by the equation (\ref{iz1}) and we defined
\begin{equation}
r_{+}(\omega)=\frac{1\,}{\sqrt{2}}\sqrt{(M-\omega){l}^{2}+l\sqrt{{(M-\omega)}%
^{2}{l}^{2}-{J}^{2}}}.  \label{iz1w}
\end{equation}
If we expand $\Delta S_{QG}$ (\ref{iz10}) and recast terms up to leading
order in $\omega $, we find
\begin{equation}
\Delta S_{QGC}\cong \Psi \omega +O(\omega ^{2}),  \label{iz11}
\end{equation}
where
\begin{eqnarray}
\Psi =&-&{\frac{l\,\left( 12\,{J}^{2}Ml+11\,{J}^{2}\sqrt{{M}^{2}{l}^{2}-{J}%
^{2}}-2\,{M}^{3}{l}^{3}-2\,{M}^{2}{l}^{2}\sqrt{{M}^{2}{l}^{2}-{J}^{2}}%
\right) \alpha }{4\sqrt{{M}^{2}{l}^{2}-{J}^{2}}\left( Ml+\sqrt{{M}^{2}{l}%
^{2}-{J}^{2}}\right) \left( {M}^{2}{l}^{2}+Ml\sqrt{{M}^{2}{l}^{2}-{J}^{2}}+{J%
}^{2}\right) }}  \notag \\
&+&{\frac{\sqrt{2l}\beta }{8\pi \,\sqrt{Ml+\sqrt{{M}^{2}{l}^{2}-{J}^{2}}}%
\sqrt{{M}^{2}{l}^{2}-{J}^{2}}}}-{\frac{\pi \,\sqrt{2}{l}^{3/2}\sqrt{Ml+\sqrt{%
{M}^{2}{l}^{2}-{J}^{2}}}}{\sqrt{{M}^{2}{l}^{2}-{J}^{2}}}}.  \label{iz12n}
\end{eqnarray}
Based on Eqs. (\ref{iz11}) and (\ref{iz12n}) and recalling the the Boltzmann
factor \cite{Refiz12}:
\begin{equation}
\Gamma _{QGC}\sim e^{\Delta S_{QGC}}=e^{-\frac{\omega }{T_{QGC}}},
\label{54}
\end{equation}
we find out the $QGC$ temperature as follows
\begin{equation}
T_{QGC}=-\frac{1}{\Psi }.  \label{55}
\end{equation}
We will show that there are a maximum mass to have positive temperature (\ref{55}). It means that this relation is only valid for small mass at quantum scales.
It is easy to verify that when suppressing the $QGC$ effects (i.e., $\alpha
=\beta =0$), $T_{QGC}$ reduces to
\begin{equation}
\left. T_{QGC}\right\vert _{\alpha =\beta =0} =\frac{\sqrt{{M}^{2}{l}^{2}-{%
J}^{2}}}{\,\sqrt{2}{l}^{3/2}\pi\sqrt{Ml+\sqrt{{M}^{2}{l}^{2}-{J}^{2}}}}
=\frac{1}{4\pi }\left( \frac{2r_{+}}{l^{2}}-\frac{J^{2}}{2r_{+}^{3}}%
\right),
\end{equation}
which is nothing but the standard Hawking temperature (\ref{iz0m}) of the
rotating BTZ black hole.

\section{GUP corrections to entropy of AO black hole}

\label{secIII}

In this section, the entropy of the AO black hole will be calculated using
the GUP corrections \cite{stringe}. At the size of Planck length,
semiclassical methods do not work properly, so that one needs to use the
theory of the quantum gravity which is not complete yet. For this purpose,
some corrections to the classical theory is used to approach quantum gravity
regime. Now, we study the GUP corrections, which were first applied in
string theory and in loop quantum gravity, on the Hawking temperature and
Bekenstein entropy \cite{Ovgun:2017hje,Sakalli:2016mnk,Ovgun:2015box,Ovgun:2015jna,Ovgun:2016roz,Alonso-Serrano:2018ycq}.
GUP can directly be applied to the modified thermodynamical quantities
by counting the number of states with the help of quantum corrections \cite{alpha1,alpha2}. Furthermore, GUP provides the modification of the
Heisenberg principle at Planck scales \cite{Ovgun:2017hje,Sakalli:2016mnk,Ovgun:2015box,Ovgun:2015jna,Ovgun:2016roz,Alonso-Serrano:2018ycq,stringe,alpha1,alpha2,GUP2,GUP4}
\begin{equation}
\Delta x\Delta p\geq \hbar \left( 1-\frac{\gamma l_{p}}{\hbar }\Delta p+%
\frac{\gamma ^{2}l_{p}^{2}}{\hbar ^{2}}(\Delta p)^{2}\right) ,  \label{gup}
\end{equation}%
where $\gamma$ is a dimensionless positive parameter, $l_{p}$ and $M_{p}$ denote the Planck length and the Plank mass respectively, also $c$ denotes the light velocity. By using the Taylor series one can rewrite the equation (\ref{gup}) as follow \cite{1510.08444},
\begin{equation}
\Delta p\geq \frac{1}{2\Delta x}\left[ 1-\frac{\gamma }{2\Delta x}+\frac{%
\gamma ^{2}}{2(\Delta x)^{2}}+\cdots \right],  \label{p}
\end{equation}
where $G=c=k_{B}=1$ is used. Therefore, $\Delta x\Delta p\geq 1$, which yields $E\Delta x\geq 1$\cite{Scardigli1}, where the standard dispersion relation $E^{2}=p^{2}+m^{2}$ is used \cite{Scardigli2}. Then, for the massless particles one can obtain $E=\Delta p\geq 1/\Delta x$ \cite{Scardigli3}. In that case GUP corrected energy $E_{GUP}$ becomes \cite{1502.00179},
\begin{equation*}
\Gamma \simeq \exp [-2\mathrm{Im}(\mathcal{I})]=\exp \left[ \frac{-4{\pi }%
E_{GUP}}{\kappa }\right].
\end{equation*}%
where surface gravity $\kappa $ is given by the equation (\ref{izn10}) and
\begin{equation}
E_{GUP}\geq E\left[ 1-\frac{\gamma }{2(\Delta x)}+\frac{\gamma ^{2}}{%
2(\Delta x)^{2}}+\cdots \right] .
\end{equation}%
Comparison with the Boltzmann factor ($e^{-E/T}$), we obtain GUP corrected temperature
of the AO black hole as
\begin{equation*}
T\leq T_{H}\left[ 1-\frac{\gamma }{2(\Delta x)}+\frac{\gamma ^{2}}{2(\Delta
x)^{2}}+\cdots \right] ^{-1},
\end{equation*}%
where the Hawking temperature of the AO black hole is given by Eq. (\ref{temp}):
\begin{equation}
T_{H}=\frac{1}{4\pi }\Big(\frac{2r_{+}}{l^{2}}-\frac{J^{2}}{2r_{+}^{3}}-%
\frac{\pi Q^{2}}{2r_{+}}\Big).  \label{temp2}
\end{equation}%
In the case of near horizon of the AO black hole, we assume $\Delta x=2r_{+}$, then one can get the GUP corrected
temperature as follows
\begin{eqnarray}
T_{GUP} &\leq &\frac{1}{4\pi }\Big(\frac{2r_{+}}{l^{2}}-\frac{J^{2}}{%
2r_{+}^{3}}-\frac{\pi Q^{2}}{2r_{+}}\Big)\left( 1-\frac{\gamma }{4r_{+}}+%
\frac{\gamma ^{2}}{8r_{+}^{2}}+\cdots \right) ^{-1}  \notag  \label{Tgup} \\
&\simeq &\frac{1}{4\pi }\Big(\frac{2r_{+}}{l^{2}}-\frac{J^{2}}{2r_{+}^{3}}-%
\frac{\pi Q^{2}}{2r_{+}}\Big)\left( 1+\frac{\gamma }{4r_{+}}-\frac{\gamma
^{2}}{8r_{+}^{2}}+\cdots \right).
\end{eqnarray}%
Inserting event horizon radius in terms of the black hole hairs and fix coefficients one can re produce corrected temperature given by the equation (\ref{55}). At this point, one can remember the laws of black hole thermodynamics to
determine the entropy of the AO black hole. In terms of the black hole entropy $S_{0}=4\pi r_{+}$, we find
\begin{equation}
S_{GUP}\leq S_{0}-\gamma \pi \ln \left( S_{0}T_{GUP}^{2}\right) -\frac{2\pi ^{2}\gamma
^{2}}{S_{0}}+\cdots .  \label{Entgup}
\end{equation}
Comparing with the entropy (\ref{Gentropy}) we find $\gamma\sim\frac{\alpha}{2}$ and $\beta=-2\pi^{2}\gamma^{2}$. Hence, we have the following expression for the temperature,
\begin{equation}
T_{GUP}=\frac{1}{2\pi }\Big(\frac{2r_{+}}{l^{2}}-\frac{J^{2}}{%
2r_{+}^{3}}-\frac{\pi Q^{2}}{2r_{+}}\Big)\Big(1+\frac{\gamma l_{p}}{\Delta x}\Big)^{-1}\left[ 1+\sqrt{1-%
\frac{4}{(1+\frac{\Delta x}{\gamma l_{p}})^{2}}}\right] ^{-1},
\label{texact}
\end{equation}%
From above, we have deduced that the maximum temperature satisfies the
following relation:
\begin{equation}
T_{GUP}\leq T_{max}=\frac{1}{4\pi }\Big(\frac{2r_{+}}{l^{2}}-\frac{J^{2}}{%
2r_{+}^{3}}-\frac{\pi Q^{2}}{2r_{+}}\Big).
\end{equation}
We will show that the temperature (\ref{texact}) coincides with the higher order quantum corrected temperature (\ref{55}) at low mass. For the massive black hole $T_{QGC}\approx T_{GUP}$ for $\alpha=0$.

\section{Conclusion and discussion}
\begin{figure}[h!]
\begin{center}
$%
\begin{array}{cccc}
\includegraphics[width=95 mm]{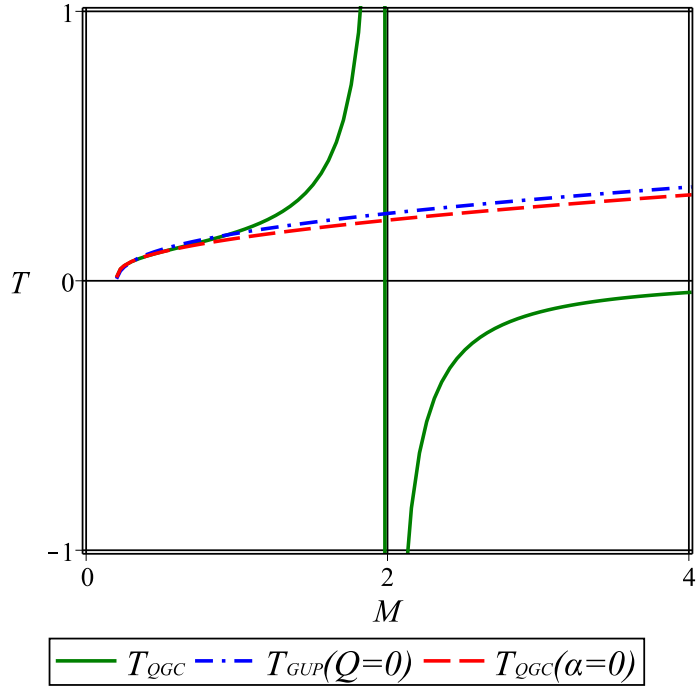}
\end{array}%
$%
\end{center}
\caption{Temperature in terms of the black hole mass for $l=1$, $J=0.2$, $\gamma=1$ and $\beta=1$. Green solid line drawn for $\alpha=1$.}
\label{fig4}
\end{figure}

In this paper, we have considered the AO black hole and study its
thermodynamics by taking into account of quantum corrections (considering the effects of back reaction and GUP separately. We have
computed the modified Helmholtz free energy and used it to investigate the $P-V$ criticality for the AO black hole. From the results obtained, we
have revealed that there are no Van der Waals behavior and critical points.
Performing the numerical analysis for the specific heat, we have shown that
the quantum corrected terms of the entropy signal the instability. We have
also derived the QGC (with back reaction effect) and GUP corrected expressions of
thermodynamic parameters: temperature, heat capacity, and entropy of
the AO black hole. In particular, for the microscopic AO black holes, those
corrections remove the thermal instability (see Fig. \ref{fig:3}). Furthermore, we have also
calculated the upper limit of the Hawking temperature and prove that $%
T_{GUP}\leq T_{max}=T_{H}$. We calculated corrected temperature due to the higher order quantum corrections and obtain $T_{QGC}$ for uncharged AO black hole. Solid green line of the Fig. \ref{fig4} represent $T_{QGC}$. Also we calculated corrected temperature due to generalized uncertainty principle and obtain $T_{GUP}$ for the uncharged AO black hole ($Q=0$) as illustrated by blue dash dotted line of the Fig. \ref{fig4}. We can see both curves coincide for the low mass ($M<1$ for $l=1$ and $J=0.2$). It is interesting to note that if we neglect $\alpha$ then, $T_{QGC}\approx T_{GUP}$ (see red dashed line of the  Fig. \ref{fig4}), which means that corrections due to generalized uncertainty principle is corresponding to higher order quantum correction.\\

To this end, we have focused on the second order
corrections that were ignored in the earlier studies about this subject.\\
The present study motivates us for doing further research in this direction. We plan
to extend our analytical analysis to the higher dimensional black holes and
explore the effects of dimension on the quantum corrected temperature and
entropy. One can also consider the Einstein-Maxwell action coupled with a charged scalar field and repeat calculations.\\
Finally it may be interesting to consider the corrections arising from the classical geometry which is called extended uncertainty principle (EUP) on the thermodynamics of black hole \cite{1809}. Following Ref. \cite{Park} in which the Hawking-Page transition of the BTZ black hole where discussed in the framework of the EUP, and its GUP corrections (GEUP), we also aim to study the EUP and GEUP corrections on the AO black hole thermodynamics.

\acknowledgments
This work is supported by Comisi\'on Nacional
de Ciencias y Tecnolog\'ia of Chile through FONDECYT Grant N$^{\textup{o}}$ 3170035 (A. \"{O}.).

\end{document}